# Asymmetric Weighted Cascade Model for Competitive Influence Maximization

Vipin Gunda*
*Cornell University*
Ithaca, New York, USA
vg245@cornell.edu

Archit Mehta*
*Cornell University*
Ithaca, New York, USA
am2283@cornell.edu

*Abstract* — We introduce a modified Weighted Cascade model that integrates asymmetric budgets and product scores, providing new insights into the Generalized Asymmetric Influence Maximization problem, which we establish as NP-hard. Our simulations demonstrate that players with higher budgets possess a distinct advantage in networks characterized by larger diameters, whereas players with superior product scores exhibit a significant advantage in networks with smaller diameters. Moreover, we identify a robust linear relationship between graph size and the magnitude of influenced nodes. In densely connected networks we derive bounds for the probabilities of influence that are independent of network size. Our examination of Nash equilibria in this domain underscores the absence of a guaranteed pure Nash equilibrium, suggesting that the strategic enhancement of budgets or product scores may yield more substantial benefits than the pursuit of an optimal strategy in this context.

## I. INTRODUCTION

The Influence Maximization problem represents a fundamental challenge in the field of networks. It aims to pinpoint the most influential nodes within a network whose activation yields the maximal spread of influence across the entire network. The realm of Competitive Influence Maximization extends this paradigm to scenarios where multiple entities vie for influence within a network. The challenge is rooted in not only identifying the ideal subset of nodes to activate, but also realistically emulating and accounting for competition. Real-world applications of Competitive Influence Maximization are abundant, spanning from marketing campaigns to political contexts to the medical context. Thus, Competitive Influence Maximization stands as an important problem that bridges theoretical network analysis to practical applications across varying domains. Within the scope of Competitive Influence Maximization, the idea of asymmetric influence introduces a layer of realistic complexity, highlighting scenarios where entities possess varying degrees of initial resources. Asymmetric Influence Maximization, therefore, considers the disparities in capabilities among entities, requiring a more complex analysis of how these inequalities affect starting strategy selection. Consequently, these nuances amplify both the realism and complexity of Competitive Influence Maximization, potentially outlining the ideal behavior for real firms based on their size or resources along with network type.

## II. RELATED WORKS

Recent advancements in Competitive Influence Maximization, such as [1] and [2], have introduced game-theoretic frameworks that account for dynamic strategy adjustments among competitors. While [1] offers a novel approach by identifying Nash Equilibria across diverse strategies, its complexity limits scalability. Meanwhile, [2] highlights the role of adoption dynamics and resource allocation efficiency, introducing metrics like the "Price of Anarchy" and the Budget Multiplier, enhancing the understanding of equilibrium outcomes in competitive social networks. Building on these foundations, our work aims to further explore and refine these dynamics to address the limitations and expand applicability to more complex competitive scenarios.

## III. ASYMMETRIC WEIGHTED CASCADE MODEL

### A. Problem Definition

We define the Asymmetric Influence Maximization problem as consisting of $n$ companies $c_1, \ldots, c_n$ such that each company has a budget $k_1, \ldots, k_n$ of initial nodes that it can attempt to influence. Moreover, each company $c_i$ has a product score $p_i \in [0, 1]$. We are also given an undirected graph $G = (V, E)$. The set of initial nodes in graph $G$ that company $c_i$ tries to influence is called the *seed set* of $c_i$ and denoted as $S_i$ where $|S_i| \leq k_i$. Seed sets across companies are not necessarily disjoint. The influence of a company spreads through the network during a given "timestep," during which currently uninfluenced nodes from the previous time step are influenced based on the modified Weighted Cascade model introduced below. We also denote the set of nodes influenced by company $c_i$ at timestep $j$ as $A_i^j$. Thus, once seed sets are chosen, the actual initial nodes a company $c_i$ begins the simulation with is $A_i^0$. If a node is only part of a single seed set $S_i$, that node is automatically part of $A_i^0$. If a node is in several seed sets $\{S_1, \ldots, S_m\}$, we construct the discrete distribution $X$, where:

$$\forall i \in \{1, \ldots, m\} \; P(X = i) = \frac{p_i}{\sum_{j=1}^m p_j}$$

We then sample from this distribution, and if we denote the result of this sampling as $k$, we place the node in the set $A_k^0$.

### B. Model

After the sets $A_i^0$ are chosen, we run all subsequent time steps of the simulation as follows. If we define the set of all

---

companies as $C = c_1, c_2, \ldots, c_n$, then in the case that we have some $C' \subseteq C, |C'| > 1$ such that all companies in $C'$ have at least one influenced node neighboring a particular node $v$ at timestep $j$, then node $v$ is influenced based on the *Asymmetric Weighted Cascade Model*:

$$p_{total} \coloneqq \sum_{c_i \in C'} p_i$$

$$e_i \coloneqq \text{number of neighbors of } v \text{ in } A_i^j$$

$$\forall c_i \in C', P(v \in A_i^{j+1}) = \frac{p_i}{p_{total}} \cdot \frac{e_i}{\sum_{k=1}^n e_k} \left(1 - \left(1 - \frac{1}{\deg(v)}\right)^{\sum_{l=1}^n e_l}\right)$$

We can represent this once again as a discrete random variable, where at timestep $j$:

$$\forall c_i \in C', P(X = i) = P(v \in A_i^{j+1})$$

and

$$P(X = 0) = 1 - \sum_{c_i \in C'} P(v \in A_i^{j+1})$$

Thus, if an uninfluenced node neighbors one or more influenced nodes at a particular time step, it will construct and sample from this discrete random variable to determine whether it becomes influenced at the next time step and which node it is influenced by. If the sampled result is 0, the node remains uninfluenced and can potentially be influenced again at a future timestep. If the result is a nonzero value $k$, the node becomes influenced and is added to $A_k^{j+1}$. Moreover, once a node has been influenced by a competitor, its status can never change.

Finally, our model also takes as input a set of $n$ strategies that each company adopts in order to select their initial seed, with the set of strategies denoted $\phi_1, \ldots, \phi_n$, and company $c_i$ choosing strategy $\phi_i$ to select its seed set. Once the seed sets are selected, the simulation continues as described above until either all nodes have been influenced or have probability 1 of remaining uninfluenced in the next timestep.

## IV. NP-Hardness of Asymmetric Cascade

We now analyze the hardness of Asymmetric Competitive Influence Maximization. We begin by determining whether a more generalized version of our proposed model is NP-Hard to see if there can be a general polynomial time algorithm to solve problems of the type we are examining. Consider the same Asymmetric Weighted Cascade problem and model defined previously. Now, instead of $\forall c_i \in C', P(v \in A_i^{j+1}) = \frac{p_i}{p_{total}} \cdot \frac{e_i}{\sum_{k=1}^n e_k} \left(1 - \left(1 - \frac{1}{\deg(v)}\right)^{\sum_{l=1}^n e_l}\right)$, we allow for an arbitrary node function $f$ that will be used at each timestep to define whether a node gets influenced, and if so, by which company. Thus, the *Generalized* Asymmetric Competitive Influence Maximization problem asks whether the player $c_1$ can select an initial set of nodes with its budget such that, in expectation, it influences more than $n$ nodes by the end of the influence cascade (with the cascade occurring as described by node function $f$). Observe that our described Asymmetric Influence Maximization problem is merely a special case of this generalized problem.

We now prove that the Generalized Asymmetric Influence Maximization problem is NP-Hard by performing a reduction from Vertex Cover. Given an instance of Vertex Cover over a graph $G$ and budget $k$, we define a instance of the Generalized Asymmetric Influence Maximization problem as follows. We define the instance over precisely the same network $G = (V, E)$, with only one player $c_1$ with budget $k_1 = k$, product score $p_1 = 1$. We also define the node function $f$ such that

$$f(\{A_1^0, \ldots, A_1^j\}, \{p_1\}, v) = X$$

where $X$ is a distribution such that

$$P(X = 0) = \begin{cases} 1 & \text{neighbors}(v) \cap A_1^0 = \emptyset \\ 0 & \text{otherwise} \end{cases}$$

$$P(X = 1) = \begin{cases} 0 & \text{neighbors}(v) \cap A_1^0 = \emptyset \\ 1 & \text{otherwise} \end{cases}$$

where $\text{neighbors}(v)$ is a function that returns the set of neighbors of a given node $v$. We then ask whether the player $c_1$ can select a set of nodes with its budget such that the expected number of influenced nodes is greater than or equal to $|V|$. We now show that the provided Vertex Cover problem is a yes instance if and only if this instance of the Generalized Asymmetric Influence Maximization problem is a yes instance, thereby showing that the Generalized Influence Maximization problem is NP-Hard and completing the proof.

Observe that if the provided instance of Vertex Cover is a yes instance, then there is a set of nodes $R \coloneqq \{v_1, \ldots, v_m\}$ of size less than or equal to $k$ in the graph such that every other node in the graph is a neighbor of one of the nodes in this set. Observe that if $c_1$ thus selects precisely the set $R$ with its budget (which is possible since $k_1 = k$), $A_1^0 = R$ since $c_1$ will be the only competitor attempting to influence these nodes and thus all of these nodes will be influenced by $c_1$. Moreover, observe that on the next time step the node function $f$ for all influenced nodes will output a distribution $X$ such that $P(X = 0) = 0$ and $P(X = 1) = 1$, since all nodes in the graph will have a neighbor in $A_1^0$, and thus all nodes in the graph will be influenced by $c_1$ on the next time step and the cascade will conclude. It then follows that there exists a set of nodes $c_1$ attempts to influence with its budget that will yield an expected number of influenced nodes equal to $|V|$, and therefore, our instance of Generalized Asymmetric Influence Maximization is a yes instance.

If the instance of Generalized Asymmetric Influence Maximization is a yes instance, there exists a set of nodes $\{v_1, \ldots, v_m\}$ of size less than or equal to $k_1 = k$ in the graph such that if $c_1$ selects these nodes, the expected number of nodes influenced by $c_1$ once the propagation concludes is greater than or equal to $|V|$. Assume, for the sake of contradiction, there exists a node $n \in V$ such that $n$ is not a neighbor of any of the nodes in $\{v_1, \ldots, v_m\}$. Observe then that no matter how many time steps we take, when we compute the node function on $n$, it will always output a distribution which has probability 1 of being 0 and probability 0 of being 1. Thus, if the number of nodes influenced by $c_1$ is less than $|V|$ in every cascade, the expected number of nodes influenced by $c_1$ must be strictly less than $|V|$. This clearly contradicts the fact that the expected number of nodes influenced by $c_1$ by choosing the set $\{v_1, \ldots, v_m\}$ is greater than or equal to $|V|$, and thus our assumption must be false. It then follows that every node in $G$ is a neighbor of some node in $\{v_1, \ldots, v_m\}$ and that the size of this set is less than or equal to $k$, and hence it follows by definition that our instance of Vertex Cover is a yes instance.

We have shown that an instance of Vertex Cover is a yes instance if and only if the corresponding instance of Generalized Asymmetric Influence Maximization as constructed by our reduction above is a yes instance. We have proven Generalized Asymmetric Influence Maximization is NP-Hard. Therefore, there is no general polynomial time algorithm to solve the general version of this problem. It is still possible, however, that for our specific model for Asymmetric Influence Maximization a polynomial time solver exists because our proposed model has a fixed node function. Although we do not have a direct proof of NP-Hardness for our problem and cannot rule out a polynomial time solver, we conjecture that the problem is still NP-Hard.

## V. MODEL SIMULATIONS

As an initial step in determining the characteristics our model, we ran simulations in various settings. We began by examining how network structure, budgets, and product scores impact the number of nodes a player influences. We constructed a simulation consisting of two players: a "product player" and a "budget player." The product player's budget is equal to $n/50$, where $n$ is the number of nodes in the graph, and its product score is equal to 1. The budget player's budget is equal to $n/10$, and its product score is equal to 0.2.

The simulation also takes a function called *GenerateGraph*, which given a number $n$ as a parameter, produces a graph of some predetermined structure consisting of $n$ nodes. It also takes a function *InitializePlayer*, which given a product score and budget score, creates a "player" to pass to the *RunModel* function. The function *RunModel*, which given a set of players and graph, randomly initializes the seed sets for each player and cascades influence via our Asymmetric Weighted Cascade model until all nodes in the graph have been influenced by some player. It produces as output the number of nodes influenced by each player once all nodes have been influenced. This simulation can be described with the following pseudo code:

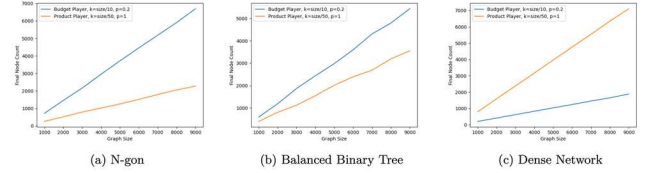

Fig. 1. Product vs Budget Simulation.

We ran this simulation on three different graph structures: an n-gon (a circular graph consisting of $n$ nodes), a balanced binary tree, and a dense network (a graph where every node is connected to every other node). Each network is an undirected graph. We graphed the number of nodes influenced by each player against the size of the graph. We see in Figure 2(a) that as the number of nodes number of nodes in the n-gon graph increases, the budget player appears to dominate the product player, with the gap between the number of nodes influenced by each player increasing. In the case of the Balanced Binary Tree, we see in Figure 2(b) that although the budget player still influences more nodes than the product player as the graph size increases, the gap between the two players is much smaller than in the n-gon. Finally, we see in Figure 2(c) that the product player dominates the budget player in the dense network.

(a) N-gon  (b) Balanced Binary Tree  (c) Dense Network

Fig. 2. Product vs Budget simulations on various network shapes. These graphs capture the final number of nodes influenced by the budget player and product player on a particular type of network at a given graph size.

Quite interestingly, there appears to be a pattern that emerges in the figures based on the diameter and node degrees of the graph type. Specifically, observe that if all three types of networks were initialized with the same number of nodes $n$ (assuming $n \geq 1000$), the n-gon would have the highest diameter of $n/2$ and lowest node degree (with every node at degree 2). The balanced binary tree would have the second highest diameter of $2 \log n$ and second lowest node degree (with every node at degree 2 or 3). The dense network would have lowest diameter of 1 and the highest node degree (with every node having degree $n - 1$). It appears as though players with higher budgets have an inherent advantage in networks where diameter is quite high (and thus influence diffuses slower), and that players with higher product scores have an inherent advantage in networks where diameter is low (and thus influences diffuses faster). This is a relationship we will examine rigorously in the next section. Moreover, we observed that the plotted series for each player in each of the graphs in Figure 2 were strongly linear, and found that for each plotted line, $R^2 \geq 0.98$.

## VI. NETWORK-SPECIFIC MODEL DERIVATIONS

We begin by examining the case of the dense network presented in the simulation above, assuming we have a product player $c_1$ with product score $p_1$ and budget $b_1$, and budget player $c_2$ with product score $p_2$ and budget $b_2$. Moreover, $b_1 < b_2$, $p_1 > p_2$, $\frac{b_1}{b_1+b_2} = \frac{p_2}{p_1+p_2}$, and $\frac{b_2}{b_1+b_2} = \frac{p_1}{p_1+p_2}$,. We begin by making the simplifying assumption that the players choose entirely distinct nodes for each of their seed sets. This assumption makes sense in our simulation because the budgets are relatively small compared to the size of the graph, and thus we likely would not expect there to be a significant amount of nodes that *both* players selected. We now examine what the first step of the influence cascade under our model looks like.

Observe that for every node $v_i$ in the network, $P(v_i \in A_1^1) = \frac{p_1}{p_1+p_2} \cdot \frac{b_1}{b_1+b_2} \cdot (1 - (1 - \frac{1}{n-1})^{b_1+b_2})$ and $P(v_i \in A_2^1) = \frac{p_2}{p_1+p_2} \cdot \frac{b_2}{b_1+b_2} \cdot (1 - (1 - \frac{1}{n-1})^{b_1+b_2})$. These equations follow directly from our model definition and since the dense network is fully connected, $deg(v) = n - 1$ for every node and every node in $A_i^0$ must be a neighbor of the node $v_i$. Since $\frac{b_1}{b_1+b_2} = \frac{p_2}{p_1+p_2}$ and $\frac{b_2}{b_1+b_2} = \frac{p_1}{p_1+p_2}$, observe that these probabilities are precisely equal to one another for every node in the graph. It follows that the expected increase in nodes for both players is equal. If we denote this expected increase in nodes as $x$, in expectation $P(v_i \in A_1^2)$ will be $\frac{p_1}{p_1+p_2} \cdot \frac{b_1+x}{b_1+b_2+2x} \cdot (1 - (1 - \frac{1}{n-1})^{b_1+b_2})$ and $P(v_i \in A_2^2)$ will be $\frac{p_2}{p_1+p_2} \cdot \frac{b_2+x}{b_1+b_2+2x} \cdot (1 - (1 - \frac{1}{n-1})^{b_1+b_2})$. By definition we had $b_2 > b_1$, and thus $\frac{b_1}{b_1+b_2} < 0.5$ and $\frac{b_2}{b_1+b_2} > 0.5$. Since $\frac{b_1}{b_1+b_2} < 0.5$, observe that for any positive nonzero number $a$, we have $\frac{b_1}{b_1+b_2} < \frac{a}{2a}$, which implies $\frac{a}{b_1} > \frac{2a}{b_1+b_2}$ and thus $1 + \frac{a}{b_1} > 1 + \frac{2a}{b_1+b_2}$, which finally implies $\frac{b_1+a}{b_1} > \frac{b_1+b_2+2a}{b_1+b_2}$. Symmetrically, we have $\frac{b_2+a}{b_2} < \frac{b_1+b_2+2a}{b_1+b_2}$. It therefore follows that $\frac{b_1+x}{b_1+b_2+2x} > \frac{b_1}{b_1+b_2}$ and also that $\frac{b_2+x}{b_1+b_2+2x} < \frac{b_2}{b_1+b_2}$. Hence, based on our equations above, we expect for all nodes $v_i$ that $P(v_i \in A_1^2) > P(v_i \in A_2^2)$.

We therefore expect that on the second time step the expected number of nodes gained by player $c_1$ will be greater than that of $c_2$, which will allow $c_1$ to continue influencing more nodes in expectation in future time steps than $c_2$, ultimately allowing the product player $c_1$ to dominate. This is precisely what we see in our simulation, and thus this derivation supports our results. Moreover, observe that if we remove our assumption that the two players do not select any of the same nodes with their budget, this only increases the advantage of the product player $c_1$, seeing as when two players attempt to select the same seed node, the tie is resolved based on product score. Then we would expect $\frac{|A_1^0|}{|A_2^0|} > \frac{b_1}{b_2}$, which implies that $P(v_i \in A_1^1) > P(v_i \in A_2^1)$, meaning that once again we expect the product player to dominate over time. We have therefore proven mathematically for the dense network that given the setting of our simulation, the product player must dominate over time.

Another question we can ask is how the various parameters of this simulation (such as the product scores and budgets of each player) impact the cascade. In other words, we can attempt to explicitly bound the probabilities $P(v_i \in A_1^1)$ and $P(v_i \in A_2^1)$ in terms of $b_1, b_2, p_1,$ and $p_2$. We know from earlier that:

$$P(v_i \in A_1^1) = \frac{p_1}{p_1+p_2} \cdot \frac{b_1}{b_1+b_2} \cdot (1 - (1 - \frac{1}{n-1})^{b_1+b_2})$$

Moreover, observe that if $b_1$ and $b_2$ were constant, it would follow trivially that as $n$ approaches infinity, the probability that a particular node is influenced becomes arbitrarily small. We therefore examine the case where $b_1 = cn$ where $c < 1$ and $b_2 = mcn$ for some constant $m > 1$. Thus $b_1 + b_2 = (m+1)cn$ and $\frac{b_1}{b_1+b_2} = \frac{cn}{(m+1)cn} = \frac{1}{m+1}$. Hence, we now have:

$$P(v_i \in A_1^1) = \frac{p_1}{p_1+p_2} \cdot \frac{1}{m+1} \cdot (1 - (1 - \frac{1}{n-1})^{(m+1)cn})$$

Now, by applying the law of multiplication of limits, we have:

$$\lim_{n \to \infty} (1 - \frac{1}{n-1})^n$$

$$= (\lim_{n \to \infty} (1 - \frac{1}{n-1})^{n-1}) \cdot (\lim_{n \to \infty} (1 - \frac{1}{n-1})^1)$$

$$= \lim_{n \to \infty} (1 - \frac{1}{n-1})^{n-1}$$

It now follows from the theorem that $\frac{1}{4} \leq (1 - \frac{1}{k})^k \leq \frac{1}{e}$ that:

$$\frac{1}{4} \leq (1 - \frac{1}{n-1})^{n-1} \leq \frac{1}{e}$$

And thus:

$$\frac{1}{4} \leq (1 - \frac{1}{n-1})^n \leq \frac{1}{e}$$

$$(\frac{1}{4})^{(m+1)c} \leq (1 - \frac{1}{n-1})^{(m+1)cn} \leq (\frac{1}{e})^{(m+1)c}$$

$$1 - (\frac{1}{4})^{(m+1)c} \geq 1 - (1 - \frac{1}{n-1})^{(m+1)cn} \geq 1 - (\frac{1}{e})^{(m+1)c}$$

$$\frac{p_1}{p_1+p_2} \cdot \frac{1}{m+1}(1 - (\frac{1}{4})^{(m+1)c}) \geq P(v_i \in A_1^1) \geq \frac{p_1}{p_1+p_2} \cdot \frac{1}{m+1}(1 - (\frac{1}{e})^{(m+1)c})$$

And we have thus found an explicit bound for $P(v_i \in A_1^1)$. Symmetrically, $P(v_i \in A_2^1)$ must then be:

$$\frac{p_2}{p_1+p_2} \cdot \frac{m}{m+1}(1 - (\frac{1}{4})^{(m+1)c}) \geq P(v_i \in A_2^1) \geq \frac{p_2}{p_1+p_2} \cdot \frac{m}{m+1}(1 - (\frac{1}{e})^{(m+1)c})$$

We have thus derived bounds for $P(v_i \in A_1^1)$ and $P(v_i \in A_2^1)$ independent of the size of the dense network. This provides further explanation as to why the data we observed in our simulation was so strongly linear: because we can impose a relatively tight bound on the probability that a particular node is influenced by a particular player on the first time step that is independent of graph size, it follows that on subsequent time steps probabilities that uninfluenced nodes become influenced by a particular player should also be independent of graph size. It then follows that the expected number of nodes influenced by the product or budget player at the end of the simulation should increase linearly with as graph size increases (since probabilities remain essentially the same across graph sizes), which is precisely the pattern we see in our simulation.

Moreover, a key result of the derivations performed on this dense network setup is that we have validated our hypothesis that the diameter and node degrees of a graph plays a significant role in determining whether the product or budget player have a significant advantage. In our first derivation above we proved that because the diameter of the dense network is constant (every node neighbors every other node in the graph), this meant that assuming the ratios of each player's budgets and product scores were equal, there would be an equal probability for every node to be influenced by the product or budget player on the first time step. Observe that in the case of the n-gon where diameter is quite high, we would observe the opposite phenomenon: most nodes would have only neighbors that were part of the budget player's seed set (since the budget player can select more nodes), and thus the expected increase in the budget player's influenced nodes on the first time step (and thus subsequent time steps) would be far greater than that of the product player, eventually allowing the budget player to dominate.

VII. NASH EQUILIBRIUM

One of the key findings in [1] was that under the setting proposed by the paper's authors, there would always exist a pure or mixed Nash equilibrium for the two players competing for influence maximization. A pure Nash Equilibrium is defined as being one where each player has a strategy that dominates all others regardless of what the opposing player selects (i.e., no player can improve their situation by changing strategies). A Mixed Nash Equilibrium is one where at least one player has a randomized strategy based on a distribution over all strategies from which they can sample from, with these distributions enabling the players to achieve an equilibrium (i.e., no player can improve their outcomes by switching strategies).

We can ask a similar question for our own model, attempting to see whether or not there always exists some Nash equilibrium. We can begin by conducting a simulation on the Hep Dataset for Academic Collaboration (consisting of roughly 10k nodes and 50k edges). In our simulation, we defined two competing players $c_1$ and $c_2$ such that $c_1$ has budget $k_1 = 500$ and product score $p_1 = 0.1$ and $c_2$ has budget $k_2 = 50$ and product score $p_2 = 1$. Each competitor can choose between three possible strategies: Single Discount, Degree Discount, and Highest Degree (where Single Discount and Degree Discount are introduced in [2] and Highest Degree corresponds to a player with budget $k$ selecting the $k$ nodes in the graph of highest degree).

We conducted simulations for every possible configuration of strategies for the players, with the results in the Game Theory matrix below, where we denote $\phi_1 :=$ Single Discount, $\phi_2 :=$ Degree Discount, $\phi_3 :=$ Highest Degree:

|  |  | Company $c_2$ | | |
|---|---|---|---|---|
|  |  | $\phi_1$ | $\phi_2$ | $\phi_3$ |
| Company $c_1$ | $\phi_1$ | (3380, 5237) | (3262, 5367) | (3359, 5252) |
|  | $\phi_2$ | (3395, 5223) | (3234, 5380) | (3156, 5466) |
|  | $\phi_3$ | (3474, 5152) | (3415, 5222) | (3680, 4940) |

Fig. 3. Game Theory Matrix for Different Strategies.

These results clearly depict that there is no pure Nash Equilibrium in this scenario, seeing as if company $c_1$ selects strategy $\phi_1$, the optimal strategy for $c_2$ is $\phi_1$, whereas if $c_1$ selects strategy $\phi_2$, the optimal strategy for $c_2$ is $\phi_3$, and thus there is no dominating strategy for player c2 and hence, there cannot exist a Pure Nash equilibrium. Thus, we have clearly shown that there does not always exist a Pure Nash equilibrium in the Asymmetric Influence Maximization problem. However, [3] has shown that there always exists a Mixed or Pure Nash equilibrium in any game with a finite set of strategies, and thus there must exist a Mixed Equilibrium. With an infinite number of options, however, it is not fully clear how players would be able to determine which distribution of strategies would yield a mixed equilibrium (especially if they are unaware of other players' product scores and budgets).

This result is quite interesting, because it suggests that rather than attempting to choose an optimal strategy, players should instead focus on either increasing their budget or improving their product scores (by, for example, adding new features to their product) based on their intuition about the network, because as we have shown above, the budget and product score of a player can give them an inherent advantage in a given network. Indeed, we can observe this phenomenon in the Game Theory matrix above as well, seeing as the player with the higher product score, $c_2$, consistently outperforms player $c_1$ on across all possible combinations of strategies for the two players.

Since the ratio of the budgets of the two players are equal to the ratio of their product scores, based on our observations above, we should therefore expect that nodes in the Hep dataset have relatively high average degree and the network itself has relatively low diameter. After computing these metrics for the Hep network, we see that this does indeed hold true, seeing as the average node in the network has diameter $\approx 5.26$, which is higher than the node degrees of the n-gon (deg 2) and balanced binary tree (deg 3), where we would expect the budget player to dominate. Similarly, the graph diameter is 18, which is smaller than the diameter in the n-gon ($\frac{|V|}{2} = 5000$) and balanced binary tree ($2 \log_2(n) \approx 26$). Hence we see that the Hep dataset is relatively densely connected, and our prediction that a player with higher product score would have an inherent advantage in such a network does indeed appear to hold.

### VIII. CONCLUSION

In summary, our contributions start with our modified version of the Weighted Cascade model that integrates asymmetric budgets and product scores. We then established the NP-hardness of the Generalized Asymmetric Influence Maximization problem through a rigorous reduction from Vertex Cover. We then presented simulations across varied graph structures, where we found that players with higher budgets have an inherent advantage when network diameter is high, and that players with higher product scores have an inherent advantage when diameter is low. We also showed that the relationship between graph size and number of nodes influenced is strongly linear throughout the simulation configurations. Furthermore, we proved why players with higher product scores would dominate in the specific case of a densely connected network and derived bounds for $P(v_i \in A_1^1)$ and $P(v_i \in A_2^1)$ independent of the size of the dense network. We used these proofs to validate our hypothesis that a graph's diameter significantly influences the advantage between higher product scores or budgets. Finally, our investigation into Nash equilibrium in asymmetric influence scenarios revealed the absence of a guaranteed pure Nash equilibrium, prompting uncertainty about how a Mixed Nash Equilibrium can be found. Consequently, our findings advocate for an emphasis on enhancing budgets or product scores over striving for an optimal strategy, thus enhancing understanding and strategic implications within asymmetric influence dynamics. Future work will focus on alternative network structures for the simulation, the development of approximation to address the NP-hardness, and deriving further Nash Equilibrium guarantees.